\documentclass [11pt]{article}
\usepackage[dvips]{graphicx}

\usepackage{epsfig}
\usepackage{graphicx}
\usepackage{inde ntfirst}
\usepackage{amsmath}
\usepackage{amsfonts}
\usepackage{amssymb}
\usepackage{hyperref}
\usepackage{latexsym}
\usepackage{color}

\begin{document}
%%%%%%%%%%%%%%%%%%%%%%%%%%%%%%%%%%%%%%%%%%%%%%%%%%%%%%%%%%%%%%%%%%%%%%%%

\begin{center}
\begin{flushright}\begin{small}    
\end{small} \end{flushright} \vspace{1.5cm}
\Large{\bf Cosmological Study of Autonomous Dynamical Systems in Modified Tele-Parallel Gravity
} 
\end{center}

\begin{center}
M. G. Ganiou $^{(a)}$\footnote{e-mail:moussiliou\_ganiou @yahoo.fr},
P. H. Logbo   $^{(b)}$ \footnote{ pascoloo@yahoo.fr}
M. J. S. Houndjo   $^{(a,b)}$\footnote{ e-mail: sthoundjo@yahoo.fr},
and J. Tossa $^{(a)}$\footnote{e-mail: joel.tassa@imsp-uac.org}
\vskip 4mm
$^a$ \,{\it Institut de Math\'{e}matiques et de Sciences Physiques (IMSP)}\\
 {\it 01 BP 613,  Porto-Novo, B\'{e}nin}\\
  $^{b}$\,{\it Facult\'e des Sciences et Techniques de Natitingou - Universit\'e de Parakou - B\'enin} \\
%$^{d}$\, {\it Ecole Normale Sup\'erieure de Natitingou - Universit\'e de Parakou - B\'enin}\\ 
\vskip 2mm
\end{center}

%%%%%%%%%%%%%%%%%%%%%%%%%%%%%%%%%%%%%%%%%%%%%%%%%%%%%%%%%%%%%%%%%%%%%%%%%%%%%%%%%%%%%%%%%%%%%%%%%%%%%%%%%%%%%%%%%%%%%%%%%%%%%%%%%%%%%%%%%%%%%%%%%%%%%%%
\begin{abstract}
\hspace{0,2cm} 
Cosmological approaches of  autonomous dynamical system in the framework of $f(T)$ gravity are investigated in this paper. Our methods applied to flat Friedmann-Robertson-Walker equations in $f(T)$ gravity, consist to extract dynamical systems whose time-dependence is contained in a single parameter $m$ depending on  the Hubble rate of Universe and its second derivative. In our attempt to investigate the autonomous aspect of the  dynamical systems reconstructed in both vacuum and non-vacuum $f(T)$ gravities, two values of the  parameter $m$  have been considered for our present analysis. In the so-called quasi-de Sitter inflationary era ($m\simeq0$),  the corresponding autonomous dynamical systems  provide  stable de Sitter attractors and unstable  de Sitter fixed points. Especially in the vacuum $f(T)$ gravity, the approximate form of the $f(T)$ gravity near the stable and the unstable de Sitter fixed points has been performed. The  matter dominated era case $(m=-\frac{9}{2})$ leads to unstable fixed points confirming matter dominated era or not, and stable attractor  fixed point describing dark energy dominated era. Another subtlety around  the stable fixed point obtained at matter dominated case in the non-vacuum $f(T)$ gravity is when the dark energy dominated era is reached, at the same time, the radiation perfect fluid dominated  succumbs.
\end{abstract}
%%%%%%%%%%%%%%%%%%%%%%%%%%%%%%%%%%%%%%%%%%%%%%%%%%%%%%%%%%%%%%%%%%%%%%%%%%%%%%%%%%%%%%%%%%%%%%%%%%%%%%%%%%%%%%%%%%%%%%%%%%%%%%%%%%%%%%%%%%%%%%%%%%%%%%%

%%%%%%%%%%%%%%%%%

%%%%%%%%%%%%%%%%%%%%%%%%%%%%%%%%%%%%%%%%%%%%%%%%%%%%%%%%%%%%%%%%%%%%%%%%%%%%%%%%%%%%%%%%%%%%%%%%%%%%%%%%%%%%%%%%%%%%%%%%%%%%%%%%%%%%%%%%%%%%%%%%%%%%%%%
\tableofcontents

\section{Introduction}
It has been supported that in addition to the inflationary stage \cite{Starobinsky} in the early Universe, the current accelerated expansion of the universe has been strongly confirmed by some independent
experiments such as the Cosmic Microwave Background Radiation (CMBR) \cite{Spergel},Type Ia Supernovae \cite{P}, large scale structure \cite{T}, baryon acoustic oscillations (BAO) \cite{E} as
well as weak lensing \cite{J}  and Sloan Digital Sky Survey (SDSS) \cite{Adelman}. In an attempt to explain this phenomenon there are two
possible paths; the first option  proposes corrections to General Relativity as the cosmological constant, the second assumption assuming that there is a dominant component of the universe, called dark 
energy. Any way that we intend to follow, there are numerous models that attempt to explain this effect,
namely, General Relativity (GR) or Teleparallel Theory Equivalent of  GR (TEGR)\cite{a'}. In the way of GR modification, one can meet the following gravitational theories:  ( $f(R)$, $f(R,\mathcal{T})$ \cite{ma1}-\cite{ma6}, $f(G)$ \cite{mj1}-\cite{mj5}),
where $R$ is the curvature scalar, $\mathcal{T}$ the trace of the energy momentum tensor, $G$ the invariant of Gauss-Bonnet defined as
$G=R^2-4R_{\mu\nu}R^{\mu\nu}+ R_{\mu\nu \lambda\sigma}R^{\mu\nu\lambda\sigma}$. Among other  additional gravitational theories to GR, the Tele-Parellel theory, based on curvatureless  Weitzenb\"{o}k connection, has been proved to be equivalent to GR \cite{equiv}. Because of its incapacity to explain some physical phenomena ( the UV modifications to GR and
also the inflation, dark energy cosmology, potential detection of gravitational waves,  etc.... ), it has been proceeded to its correction leading to the so-called $f(T)$ gravity  [\cite{st1}-\cite{st41}], and its extension $f(T,\mathcal{T})$ and $f(T, T_G)$ \cite{ok} where $T$ denotes the torsion scalar and $T_G$ the Tele-Parallel Equivalent of the Gauss-Bonnet combination.   \par 
In the present work, the $f(T)$ gravity is particularly addressed with the main goal of studying dynamical system susceptible to reproduce   some available  cosmological features. Such study is not performed for the first time in the framework of gravitational modified theories in general [\cite{dy1}-\cite{Sergy}] and in  $f(T)$ gravity in particular \cite{dy5}-\cite{Mubasher}. Indeed, it is investigated in \cite{Mubasher} the interacting dark energy model in $f(T)$ cosmology. By assuming dark energy as a perfect fluid and choosing a specific cosmologically viable $f(T)$  form, the authors of this work have  shown that there is one attractor solution to the dynamical equations of $f(T)$ Friedmann equations. They have also gotten   critical points from which arises the coupling between the  dark energy and matter.  Moreover, the authors in \cite{dy5}  have explored some aspects of  autonomous system   in the nonlocal $f(T )$ gravity from cosmological   equations describing the whole evolution history of the Universe. By building a dynamic system taking into account the interaction between matter, dark energy, radiation and a scalar field, they have demonstrated that for nonlocal $f(T )$ gravity, one can obtain the stable de Sitter solutions even in vacuum spacetime. \par 
 Here, in aim to reveal more cosmological features of our Universe evolution,  we focus our attention on  autonomous dynamical system case  through  $f(T)$ gravity in the presence, or not, of perfect matter fluids. Let's also emphasis that, extracting autonomous dynamical system in modified theory of gravity is not a trivial investigation. It  requires a rigorous choice not only for the equation of motion but also for the cosmological variables in order to find some results in agreement with the actual  phenomenological tendencies in cosmological physics and astrophysics. Why studying autonomous dynamical system? 
 The motivation for studying the autonomous limit of the cosmological dynamical system, comes mainly from the fact that if a dynamical
 system is non-autonomous, then the stability of the fixed points is not guaranteed by using the theorems which hold
 true in the autonomous case. Hence, the stability of a fixed point corresponding to a non-autonomous system is
 rather a problem without a solution, unless highly non-trivial techniques are employed. In \cite{Sergy}, S. Odintsov and its collaborator performed a detailed analysis of the $f(R)$ gravity phase space, by using an autonomous dynamical system approach. The autonomous  property of their dynamical system  is reached when  the single parameter $m$ of the system, depending on Hubble rate and its second derivative order, is constant. Two different values of $m$ are considered; the first, $m\simeq0$, corresponds to quasi-de Sitter inflationary era and makes true the slow-roll condition. The second, $m=-\frac{9}{2}$, stays for matter dominated era. Through vacuum and non-vacuum  
 $f(R)$ gravities, various fixed points with very interesting physical significance have been found and their stability has also been performed.\par
  This  very interesting previous work has been confronted to the existence of non physical fixed points (complex coordinates fixed points) which does not allow a deeply cosmological interpretations. Furthermore, the $f(T)$ theory of gravity is not equivalent to $f(R)$ theory  even mathematically, although the geometrical equivalence between Tele-Parallel theory and General Relativity. An palpable example can be seen in \cite{Bamba} where the trace-anomaly driven inflation related to $T^2$ model and those related to $R^2$ do not lead to the same inflationary scenario: $T^2$ produces de Sitter inflation with graceful against quasi-de Sitter inflation  with graceful in the case of $R^2$ model. Nevertheless, it doesn't exclude the case where both theories can coincide in the physical interpretation of a same mathematical problem or two different mathematical problems. Considering these previous remarks, we follow the same approach in \cite{Sergy} to investigate  the autonomous dynamical system in the framework of $f(T)$ theory in order to provide maximum real points with physical significance.\\
  The present paper in organised as follow. In section Sec.$2$, after giving the essential basic notions in Tele-Parallel gravity, we establish the fundamental $f(T)$ equation of motion valid for Friedmann-Robertson-Walker (FRW) space-time. These equations in vacuum case, in particular the second will be used in section Sec.$3$ to extract the autonomous dynamical corresponding to  the vacuum $f(T)$ gravity. We determinate the fixed points and analyse numerically the behaviours of dynamical system solutions for  each value of the parameter $m$. Physical interpretations around each fixed point is given by taking into consideration  the parameter of equation of state. The same study is done in the section Sec.$4$  where the radiation perfect fluid is added to the $f(T)$ gravity. We conclude our work  in the last section Sec.$5$.

%%%%%%%%%%%%%%%%%%%%%%%%%%%%%%%%%%%%%%%%%%%%%%%%%%%%%%%%%%%%%%%%%%%%%%%%%%%%%%%%%%%%%%%%%%%%%%%%%%%%%%%%%%%%%%%%%% 
\section{Motion equations in FRW Space-time in the framework of $f(T)$ gravity }
Before finding out the  Friedmann-Roberson-Walker motion equations in this work, let's present here  the essential of  fundamental notions in Tele-Parallel gravity, source  of  $f(T)$ theory. In general,  
when formulating theories of gravity, the metric tensor is of paramount importance. 
It contains the information needed to locally measure distances and thus
to make theoretical predictions about experimental findings. Furthermore, the structure of the spacetime can be described by   an
alternative dynamical variable, the well known  non-trivial tetrad $h^a_{\;\;\mu}$ which is a set of four
vectors defining a local frame at every point. The tetrads represent the   basic entity of the theory of
Teleparallel gravity. From their reconstruction  arises the  Teleparall theory as gravitational theory naturally based on the
gauge approach of the group of translations. The tretrads are  defined from  the gauge covariant derivative
for a scalar field as $h^a_{\;\;\mu}= \partial_\mu x^a+A^a_{\;\;\mu}$ with $A^a_{\;\;\mu}$ the translational gauge potential and $x^a$
the tangent-space coordinates. The tretrad $h^a_{\;\;\mu}$ and its inverse $h_a^{\;\;\mu}$ satisfy the following relations   
\begin{equation}
h^a_{\;\;\mu}h_a^{\;\;\nu}=\delta^\nu_\mu \quad \quad h^a_{\;\;\mu}h_b^{\;\;\mu}=\delta^a_b. 
\end{equation}
The Weitzenb\"{o}ck  connection \cite{Hayashi}  constitutes 
 the fundamental connection of the theory and can expressed via  
 \begin{equation}
 \Gamma_{\mu\nu}^{\lambda}= h_a^{\;\;\lambda}\partial_\nu h^a_{\;\;\mu}
 =-h^a_{\;\;\mu}\partial_\nu h_a^{\;\;\lambda}.
 \end{equation}  
 Generally,
  Latin alphabet $(a, b, c, ... = 0, 1, 2, 3)$ is used to denote the tangent space
  indices where as the Greek alphabet $(\mu,\nu, \rho, ... = 0, 1, 2, 3)$ stays for the spacetime
  indices. The space-time and its tangent space are related by their metric via  
  \begin{equation}
  g_{\mu\nu}= \eta_{ab}h^a_{\;\;\mu}h^b_{\;\;\nu},
  \end{equation}
  where  $\eta_{ab}=\text{diag}(+1,-1,-1,-1)$ is the Minkowski metric of the tangent space. Contrarily to the Levi-Civita connection, the Weitzenb\"{o}ck  connection preserve the torsion tensor whose non-vanishing component are given by 
  \begin{equation}
  T_{\;\;\;\mu\nu}^{\lambda}= \Gamma_{\;\;\;\nu\mu}^{\lambda}-
  \Gamma_{\;\;\;\mu\nu}^{\lambda}=h_a^{\;\;\lambda}(\partial_\mu h^a_{\;\;\nu}-\partial_\nu h^a_{\;\;\mu})\neq 0.
  \end{equation}
  Then, the geometrical difference between these two connections are given by the   contortion tensor $  K_{\;\;\;\mu\nu}^{\;\,\lambda}$ expressed as 
 \cite{equiv}.
  	\begin{equation}\label{cont}
  	K_{\;\;\;\mu\nu}^{\lambda}:= \Gamma_{\;\;\;\mu\nu}^{\lambda}-
  	\tilde{\Gamma}_{\;\;\;\mu\nu}^{\lambda}=\frac{1}{2}\Big (T_{\nu}{}^{\lambda}{}_{\mu}+
  	T_{\mu}{}^{\lambda}{}_{\nu}- T^{\lambda}{}_{\mu\nu}\Big),
  	\end{equation}
  where $ \tilde{\Gamma}_{\;\;\;\mu\nu}^{\;\,\lambda}$  are the Christoffel symbols or the coefficient of Levi-Civita connection.
  Furthermore the scalar torsion, fundamental element of Tele-Parallel density geometrical is given by 
  \begin{equation}\label{de}
  T:=S_\beta{}^{\mu\nu}
  T{}^\beta{}_{\mu\nu},
  \end{equation}
  with  $S_\beta{}^{\mu\nu}$, specifically defined by \cite{Hayashi}
  \begin{equation}\label{sup}
  S_\beta{}^{\mu\nu}=\frac{1}{2}\Big(  
  K{}^{\mu\nu}{}_{\beta}+
  \delta^\mu_\beta T{}^{\alpha\nu}{}_{\alpha}- \delta^\nu_\beta T{}^{\alpha\mu}{}_{\alpha}  \Big).
  \end{equation} 
  From the use of the relations (\ref{cont}) and (\ref{sup}) occurs the following fundamental relations which show the equivalence between the Tele-Parallel theory and the General Relativity (see \cite{Di},\cite{Li} for details ) 
  \begin{equation}
  R=-T-2\nabla^\mu T^\nu{}_{\mu\nu},
    \end{equation}
  \begin{equation}\label{eins}
  G_{\mu\nu}-\frac{1}{2} g_{\mu\nu} T=-\nabla^\rho S_{\nu\rho\mu}-S^{\sigma\rho}{}_{\mu}K_{\rho\sigma\nu},
   \end{equation}
   where $R$ is the Ricci  scalar and $ G_{\mu\nu}$ the Einstein tensor.\\
  The action of the modified versions of  TEGR (Tele-Parallel equivalent of General Relativity \cite{So}  
  is obtained by substituting the scalar torsion in Tele-Parallel geometrical  Lagrangian density  
   by an arbitrary function of scalar torsion   obtaining modified theory $f(T)$. 
  This approach is similar in spirit to the generalization of 
  Ricci scalar curvature of Einstein-Hilbert action  by a function of this scalar  leading  to the well known 
  $ F(R) $ theory.  The  $f(T)$ theory  can be governed   by the following action
  \begin{equation}\label{act2}
  S=\frac{1}{\kappa^2}\int d^4x hf(T)+ \int d^4x h\mathcal{L}_M,
  \end{equation}
  where $h=|\text{det}(h^a{}_\mu)|$ is equivalent to 
  $\sqrt{-g}$ in  General Relativity, $\kappa^2=\frac{16\pi G}{c^4}$,  $\mathcal{L}_M$ is the Lagrangian of the matter field.
   Then, the variation of this action with respect to tetrad $h_a{}^{\mu}$ gives 
   ( \cite{Ulhoa})
   \begin{eqnarray}\label{mot}
   \frac{1}{h}\partial_\mu(hS_a{}^{\mu\nu})f_T(T)- h_a{}^{\lambda}T^\rho{}_{\mu\lambda}S_\rho{}^{\mu\nu}f_T(T) 
   +S_a{}^{\mu\nu}\partial_\mu(T)f_{TT}(T)+\frac{1}{4h} h_a{}^{\nu}f(T)= \frac {1}{4\kappa^2}T^\nu_a, 
   \end{eqnarray}
   with $f_T(T)=df(T)/dT$,   $f_{TT}(T)=d^2f(T)/dT^2$ and $T^\nu_a$, the energy-momentum tensor. 
   Furthermore, by  using the relation (\ref{eins}), the motion equation (\ref{mot}) can be recast in the following form \cite{Di}
    \begin{eqnarray}\label{mot1}
    G_{\mu\nu}=\frac{1}{f_T(T)}\Bigg[T_{\mu\nu}+\frac{1}{2}\Big(Tf_T(T)-f(T)\Big)g_{\mu\nu} -S_{\nu\mu}{}^\sigma\nabla_\sigma T f_{TT}(T)\Bigg].
   \end{eqnarray}
   Now, we are searching for the expressions of these  motion equations in the Friedmann-Roberson-Walker space-time. Such a space-time is described
    by the following line element
   \begin{equation}\label{s3}
   ds^2=-dt^2+a^2(t)\sum_{i=1,2,3} (dx^i)^2.
   \end{equation}
   Here, $a(t)$ is the scale factor and $H\equiv \dot{a}/a$ is the Hubble parameter. From (\ref{s3}), one obtains  the torsion scalar in function
   of  $H$ by $T=-6H^2$. The dot $(.)$ means here the derivative with respect to comic time $t$. Consequently, the $f(T)$ motion equations (\ref{mot})
   or   (\ref{mot1}) become
    \begin{eqnarray}
    \label{e1}
    \frac{1}{2}f(T)-Tf_T(T)+\kappa^2\rho=0, \\
    \label{e2}
    \frac{1}{2}f(T)+(6H^2+2\dot{H})f_T(T)-24H^2\dot{H}f_{TT}(T)-\kappa^2P=0, 
    \end{eqnarray}
   where $\rho$ and $P$ are respectively global energy density and the pressure of Univers content. \\
    Otherwise, from the motion equations, particularly its form presented in (\ref{mot1}), one can extract the effective parameter of the   equation of state (EoS) valid for $f(T)$ gravity
    \begin{equation}\label{etat}
    \omega_{eff}=-1-\frac{2\dot{H}}{3H^2}.
    \end{equation}
   One has now sufficient informations to extract and to analysis some $f(T)$ autonomous dynamical systems in the following sections.

% % % % % % % % % % % % % % % % % % % % % % % % % % % % % % % % % % % % % % % % % % % % % % % % % % % % % % % % % % % % % % % % % % % % % % % % 
% % % % % % % % % % % % % % % % % % % % % % % % % % % % % % % % % % % % % % % % % % % % % % % % % %
 % % % % % % % % % % % % % % % % % % % % 
\section{ Extraction and Analysis  of Autonomous Dynamical System in Vacuum $f(T)$  gravity}
\subsection{Building  Autonomous Dynamical System in Vacuum $f(T)$  gravity}
In the absence of matter  and radiation sources, namely in vacuum, the previous motion equations become :
\begin{eqnarray}
 \label{e3}
 \frac{1}{2}f(T)-Tf^{\prime}(T)=0, \\
 \label{e4}
 \frac{1}{2}f(T)+(6H^2+2\dot{H})f^{\prime}(T)-24H^2\dot{H}f^{\prime \prime}(T)=0. 
 \end{eqnarray}	
In order to extract an autonomous dynamical system with no bad number of dynamical variables, we consider the equation (\ref{e2}), from which, we pose the following dynamical variables:
\begin{eqnarray}\label{pose}
x=-\frac{\dot{F}(T)}{HF(T)},\qquad y=\frac{f(T)}{4H^2F(T)}, \qquad z=\frac{\Re}{H^2}.
 \end{eqnarray}	
Here, $F(T)=df(T)/dt$  and $\Re=3H^2+\dot{H}$. Instead of cosmic time, one can make using of e-folding number. This possibility is assured by the  following relation 
existing  between the operator derivative with respect to e-folding number and those with respect to cosmic time 
\begin{equation}
	\frac{d}{dN}=\frac{1}{H}\frac{d}{dt}.
	\end{equation}
Then, following the same approach as \cite{Sergy}-\cite{Atur} and by making using  the equation  (\ref{e2}), one can derive  the variables $x$, $y$, and $z$ with respect to e-folding number and obtain the following dynamical system:
\begin{eqnarray}\label{syst1}
	\frac{dx}{dN}&=& -m - 9 + 6 z + 3 y + 3 x - 2 zx - yx - z^2 + x^2 + y^2\\
		\frac{dy}{dN}&=& 9-3z-yz+6y+y^2\nonumber\\
			\frac{dz}{dN}&=& -m-18+12z-2z^2 \nonumber
\end{eqnarray} 
The parameter $m$ appeared in this system is defined by \cite{Sergy} 
\begin{equation}\label{para}
	m=-\frac{\ddot{H}}{H^3}
\end{equation}
According to the fact that the Hubble rate $H$ can explicitly defend  from the e-folding number or the cosmic time, the parameter $m$ is  function of e-folding number. So, it  will constraint the system (\ref{syst1}) to be explicit function of  e-folding number. Here, we  assume that  the parameter $m$  takes constant values: the corresponding dynamical system is called autonomous dynamical system. \\
One can also make using of the relation $\frac{\dot{H}}{H^2}=z-3 $ (very used in the building of the system (\ref{syst1})) to rewrite $\omega_{eff}$  in    (\ref{etat})                as 
\begin{equation}\label{ome}
\omega_{eff}=1-\frac{2}{3}z.
\end{equation}
 For the two well chosen values  of the parameter,  both dynamical system and the parameter of  equation of state will be used  to analyse the structure of  the vacuum $f(T)$ gravity  phase space.

\subsection{Analysis of the obtained dynamical system for different valors of the system parameter $m$ }
                             \begin{center}
        {\bf a-Some considerations and clarifications on dynamical system}                     	
                             \end{center}

According to its form, one can conclude that the system presented in (\ref{syst1}) is  not linear.  Its eventual time dependence  is strongly hidden by the parameter $m$. Otherwise, we have said in the previous section that if the parameter $m$ is constant, the dynamical system becomes
 autonomous. This is physically possible according to some cosmological evolution like the de Sitter and the quasi de Sitter evolutions.   Indeed, in the quasi de Sitter evolution the scalar factor is given by
 \begin{equation}\label{scal}
 a(t)= e^{H_0t-H_it^2}\qquad \text{so}, \qquad H(t)=\frac{\dot{a}}{a}=H_0-2H_it\qquad\Longrightarrow\qquad \ddot{H}=0,
  \end{equation} 
Consequently, regarding the relation (\ref{para}), the parameter $m$ corresponding to the Hubble rate of (\ref{scal}) is constant and equals to $0$. Inversely, by fixing $m$ to zero, the obtained differential equation gives $ H(t)=H_0-2H_it$ which corresponds to quasi de Sitter evolution for which the  following slow-roll conditions are satisfied 
\begin{equation}\label{slow}
H\dot{H}\gg\ddot{H} \qquad \text{and} \qquad \dot{H}\ll H^2.
\end{equation}
One can also recover the inflationary era with this value of $m$.\par 
In the continuation of this work, our investigation will involve two values of the parameter $m$ which have physical significance \cite{Sergy}: $m=0$ (quasi de Sitter case) and $m=-9/2$ (matter dominated) case. Before proceeding to the analysis of the  dynamical system corresponded to these values of the parameter $m$, we are going to present not only  the classical approach namely the linearisation procedure which is applied to dynamical system with hyperbolic fixed point but also  some essential features of the stability theory for dynamical systems. These clarifications on theory of stability have already be done in \cite{Sergy}  and based on interesting mathematical work of \cite{Stephen}.\par
Generally, the theory of stability of dynamical system lands concretely the stability of  solutions-trajectories. This stability   unveils the behaviour of the solutions and trajectories when  the initial conditions vary weakly.  Furthermore, when studying  dynamical system, it is very important  not to say unavoidable to know  the asymptotic behaviour of the solutions and trajectories, in particular the asymptotic  meaning  after a long period of time. The length and the determination of this long period of time is generally determined by the
physical scales of the theory. In this work, involving inflationary dynamical system, the asymptotic behaviour corresponds to $N \simeq 60$. e-foldings. How can us get the nature of the dynamical fixed point? \par 
In  the first hand,  the simplest and most precious behaviour   of the solution-trajectories is presented by the
 stationary points of the dynamic system in progress, as well as by periodic orbits. The different behaviour of the solutions and  especially of the trajectories around the stationary points are strongly tributary of the nature of these points  and sometimes depends from the initial condition  (see \cite{Sergy} for details). Indeed,  a fixed point can be attractor of trajectory (stable fixed point ) or asymptotic attractor  (asymptotically stable fixed point), if the trajectories are always attracted or asymptotically  attracted respectively by this point. However, the trajectories can be repulsed at the level of one fixed point. At this  moment, the fixed point is  qualified unstable fixed point. The fixed points provide a characteristic picture of the structural stability of the dynamic system. \par 
In the second hand,  a quantitative approach to appreciate the nature of the fixed  point, resides in the application of the Hartman-Grobman theorem.  One looks for the matrix of linearisation of the autonomous dynamical system and according to the algebraic nature of the eigenvalues of the linearisation matrix at  a given  point fixed, one will be able to deduct the nature of the point. Indeed,  if the eigenvalue  of the matrix of linearisation are negative real numbers, or even  complex numbers with negative real part, then the fixed point is stable (attractor).   
 If none of the eigenvalues
 are purely imaginary, or equal to zero, then the fixed point can be an attractor or a repeller (see \cite{Stephen} for details).\par
 The Hartman-Grobman linearisation theorem provides a powerful technique to study the local stability and the
 portrait of the phase space, when we have a set of hyperbolic fixed points which means that only in the case that
 the eigenvalues of the linearisation matrix have non-zero real parts.  Let $\vec{X}  \in \mathbf{R}^n $ be a non trivial solution
 to the following system of first order differential equations, called flow, 
\begin{equation}\label{matr}
	\frac{d\vec{X}}{d\lambda}=g(\vec{X}),
\end{equation}
here $g(\vec{X})$ is a locally Lipschitz, one-to one continuous map $g: \mathbf{R}^n\rightarrow \mathbf{R}^n$. 
 Let $\vec{X}_*$ denotes the location of the fixed
 points of the dynamical system (\ref{matr}), and the corresponding Jacobian matrix, which we denote as $\mathcal{J}(g)$, is equal to
\begin{equation}\label{matri}
(\mathcal{J})_{ij}=\Big[\frac{\partial g_i}{\partial X_j}\Big].
\end{equation}
In order to have stable fixed points for system (\ref{matr}), it is enough to set all eigenvalues of the Jacobian matrix so that
$\eta_i$ satisfies $Re(\eta_i)\neq0$. The Hartman theorem predicts the existence of a homeomorphism $\mathcal{F}:U\rightarrow \mathbf{R}^n$
where $U$ is an open neighborhood of  $\vec{X}_*$, such that $\mathcal{F}(\vec{X}_*)$. From this  homeomorphism  arises the so-called flow of  homeomorphism  defined by 
\begin{equation}\label{bien}
\frac{dh(u)}{d\lambda}= \mathcal{J}h(u).
\end{equation}
It is proved that $(\ref{bien})$ is a topologically conjugate flow to the dynamical system defined in $(\ref{matr})$
                         \begin{center}
                     	{\bf b- de Sitter Inflationary Attractors and their Stability in vacuum $f(T)$ for $m=0$}                     	
                          \end{center}
We begin this present analysis by the dynamical description of the quasi de sitter evolution which corresponds to $m=0$.  According to this valor, 
  the Jacobian matrix of (\ref{matri})  which linearizes the corresponding autonomous dynamical system $(\ref{syst1})$ is 
 \begin{eqnarray}
\mathcal{J}=\Bigg[\begin{array}{ccc}
3 - y - 2 z + 2 x & 3 - x + 2 y & 6 - 2 x - 2 z \\ 
0 & -z + 6 + 2 y & -3 - y \\ 
0 & 0 & 12 - 4 z
\end{array} \Bigg],
 \end{eqnarray}
where  the source functions $g_i$ are  
\begin{eqnarray*}
g_1&=&  - 9 + 6 z + 3 y + 3 x - 2 zx - yx - z^2 + x^2 + y^2,\\
g_2&=& 9-3z-yz+6y+y^2,\\
g_3&=& -18+12z-2z^2,
\end{eqnarray*} 
By solving the system of equations $g_i=0$ for general $m$ , one obtains the fixed points of the dynamical autonomous system $(\ref{syst1})$

\begin{tabular}{|c|c|c|c|c|c|}
	\hline Critical points  & $x$ & $y$ & $z$ \\ 
	\hline $X^1_*$ & $ \frac{1}{4} \left(6-\sqrt{36-12 \sqrt{2} \sqrt{-m}-2 m}-3 \sqrt{2} \sqrt{-m}\right)$  & $-\frac{\sqrt{-m}}{\sqrt{2}} $ & $\frac{1}{2} \left(6-\sqrt{2} \sqrt{-m}\right)$  \\ 
	\hline $X^2_*$ & $ \frac{1}{4} \left(6+\sqrt{36-12 \sqrt{2} \sqrt{-m}-2 m}-3 \sqrt{2} \sqrt{-m}\right)$ & $-\frac{\sqrt{-m}}{\sqrt{2}}$  & $ \frac{1}{2} \left(6-\sqrt{2} \sqrt{-m}\right)$ \\ 
	\hline$X^3_*$  & $  -\frac{\sqrt{-m}}{\sqrt{2}}$  & $-3$ & $\frac{1}{2} \left(6-\sqrt{2} \sqrt{-m}\right)$  \\ 
	\hline$X^4_*$  & $\frac{1}{4} \left(6-\sqrt{36+12 \sqrt{2} \sqrt{-m}-2 m}+3 \sqrt{2} \sqrt{-m}\right)$ & $\frac{\sqrt{-m}}{\sqrt{2}}$  & 
	$\frac{1}{2} \left(6+\sqrt{2} \sqrt{-m}\right)$ \\ 
     \hline $X^5_*$  &  $\frac{1}{4} \left(6+\sqrt{36+12 \sqrt{2} \sqrt{-m}-2 m}+3 \sqrt{2} \sqrt{-m}\right)$ & $\frac{\sqrt{-m}}{\sqrt{2}}$  & $\frac{1}{2} \left(6+\sqrt{2} \sqrt{-m}\right)$ \\ 
    \hline$X^6_*$  & $\frac{\sqrt{-m}}{\sqrt{2}}$  & $-3$ &  $\frac{1}{2} \left(6+\sqrt{2} \sqrt{-m}\right)$\\ 
	\hline 
\end{tabular}

\vspace{1cm}
Now,  one can get these points for $m=0$. For this value of $m$, we are going to take advantage of the opportunity to look for the   eigenvalues $\eta_i$ of the Jacobi matrix associated to every critical point. The results are presented in the following  table\\
\begin{center}
\begin{tabular}{|c|c|c|c|c|c|c|}
	\hline Fixed points & $x$  & $y$ & $z$ & $\eta_1$ &  $\eta_2$ &  $\eta_3$  \\ 
	\hline $X^1_*$ & $0$ & $0$ & $3$ & $0$ & $-3$ & $3$ \\ 
	\hline $X^2_*$  & $3$ & $0$ & $3$ & $0$ & $3$ & $3$ \\ 
	\hline $X^3_*$ & $0$ & $-3$ & $3$ & $0$ & $0$ & $-3$ \\ 
	\hline 
\end{tabular} 
\end{center}
\vspace{1cm}
From this table, one can see that Jacobian matrix has no eigenvalues on the unit circle at all fixed points, so all of them are non-hyperbolic. Furthermore, the fixed point   $X^3_*$ is stable where as  $X^1_*$ and  $X^2_*$ are unstable. Another important conclusion arisen from this last table is that for three critical points, one has $z=3$. By  substituting $z=3$ in (\ref{ome}), one has $\omega_{eff}=-1$ which corresponds to the de Sitter equilibria. \par
 In order to get the analytic solutions of the dynamical system $\ref{syst1}$ for $m=0$, we  try to solve  the different differential equations that it contains. Let emphasis here that, outside of the third equation of the system, the direct  analytic resolution of the system is not easy but after some transformation, one gets the following system of the solutions     
\begin{eqnarray}
\label{caf}
z(N)&=& \frac{1+6 N-3 C_1}{2 N-C_1},\\
y(N)&=&  -\frac{\frac{\sqrt{3}}{2^{\frac{3}{4}} \sqrt{3 N-\frac{3 C_1}{2}}}+3\times 2^{\frac{1}{4}} e^{(3 N-\frac{3 C_1}{2})} C_2+\frac{e^{(3
			N-\frac{3 C_1}{2})} \sqrt{3 \pi } \text{Erf}\left[\sqrt{3 N-\frac{3 C_1}{2}}\right]}{2^{\frac{3}{4}}}}{2^{\frac{1}{4}} e^{(3 N-\frac{3 C_1}{2})} C_2+\frac{e^{(3
			N-\frac{3 C_1}{2})} \sqrt{\frac{\pi }{3}} \text{Erf}\left[\sqrt{3 N-\frac{3 C_1}{2}}\right]}{2^{\frac{3}{4}}}},\\
x(N)&=& \frac{1+6 N-3 C_1}{2 N-C_1} -3-\frac{3 e^{(-3 N+\frac{3C_1}{2}})}{\sqrt{N-\frac{C_1}{2}} \left(6 C_2+\sqrt{3 \pi } \text{Erf}\left[\sqrt{3 N-\frac{3 C_1}{2}}\right]\right)}+C_3.
\end{eqnarray}
Here, $C_1$, $C_2$ and $C_3$ are constants of integration and must depend from the initial conditions where as $\text{Erf}[x]$ means  
 the error function.  From these expression, we note that for large value of e-folding number $N$, $z(N)$  stretches toward $3$, which is exactly the behaviour we indicated earlier.\par 
 In the goal to take out again the impact of the initial conditions on the behaviour of the system, we proceed to a numeric resolution of the system. The results descended of this approach should be in agreement with those gotten analytically for suitable initial conditions.
In order to illustrate the asymptotic behaviour of the dynamical system, we choose the e-folding number  as $N\in [0,60]$. We also start our cosmological numerical  analysis with the following initial conditions $x(0)=-7$; $y(0)=-3$; $z(0)=5$.
\begin{figure}[h]
	\centering
	\begin{tabular}{rll}
		\includegraphics[width=5cm, height=5cm]{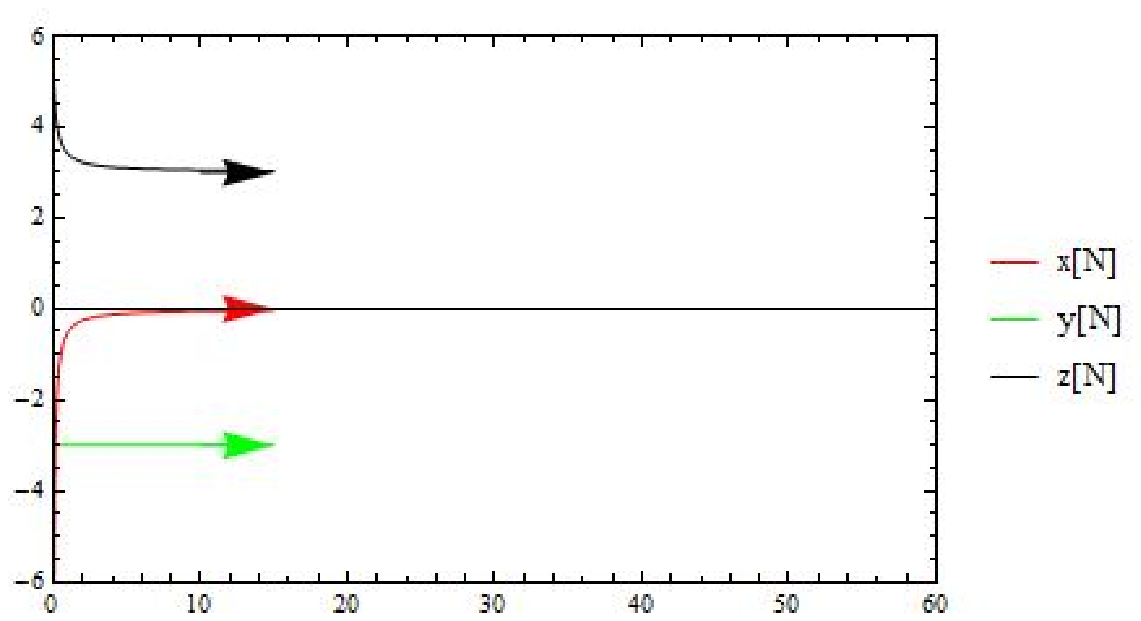}&
		\includegraphics[width=5cm, height=5cm]{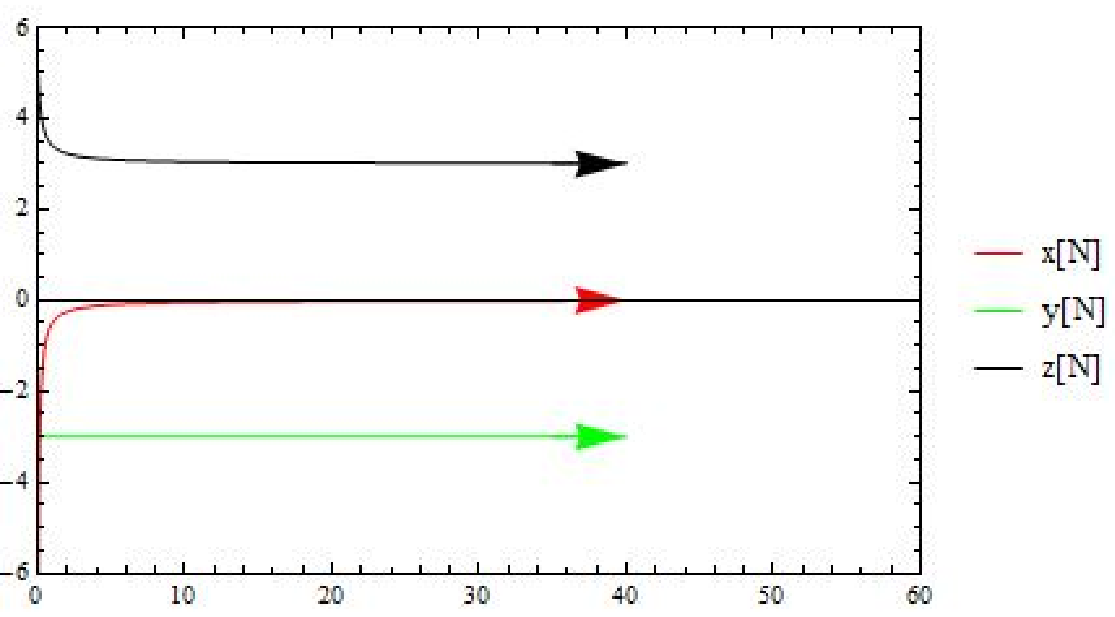}&
		\includegraphics[width=5cm, height=5cm]{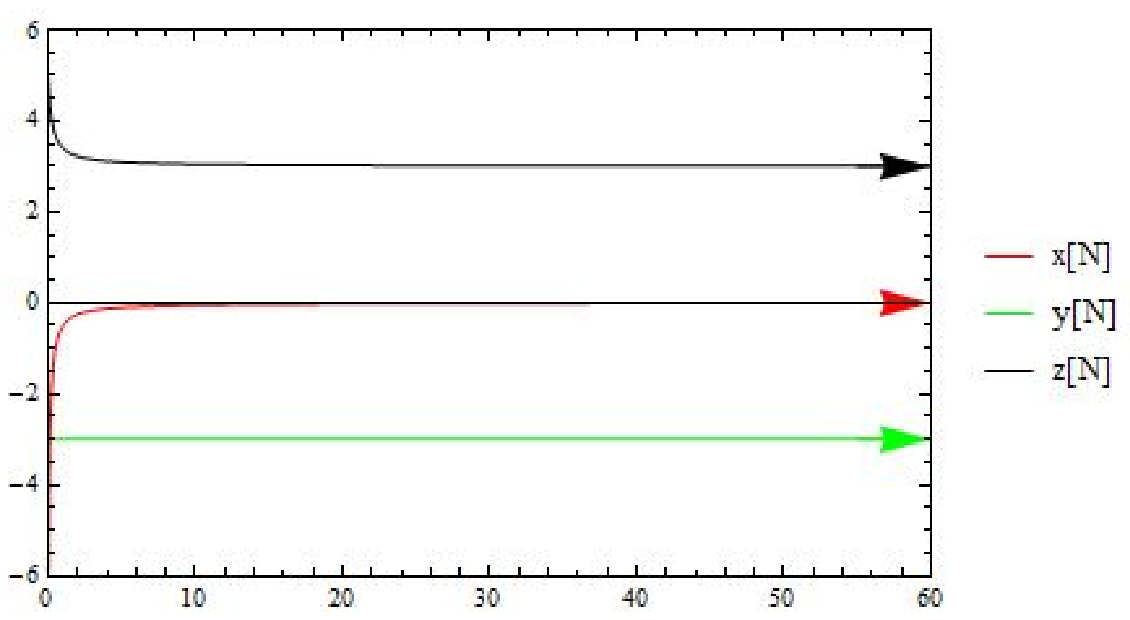}
	\end{tabular}
	\caption{  Numerical solutions $x(N)$, $y(N)$, and $z(N)$ of dynamical system (\ref{syst1})  for $m=0$ in three intervals $N\in [0,15]$, $N\in [0,40]$ and $N\in [0,60]$ for initial conditions $x(0)=-7$; $y(0)=-3$; $z(0)=5$ }
	\label{fig1}
\end{figure}
\newline
The three plots of this figure \ref{fig1} show how the  equilibrium  $X^3_*= (0 , -3,3)$ is reached progressively. The first plot (the  extreme right one) shows clearly the solution evolutions from the initial conditions and proves that the  equilibrium  $X^3_*$ is not directly reached. The two other invoke the  dreadful convergence of the solutions toward this equilibrium  $X^3_*$ and their stabilization in this last. To show more deeply, the previous properties the attractive nature of the fixed point $X^3_*= (0 , -3,3)$, we represent the solutions in the plan $x-z$  for three different initial conditions through the plots of the figure \ref{fig2}.
\begin{figure}[h]
	\centering
	\begin{tabular}{rll}
		\includegraphics[width=5cm, height=5cm]{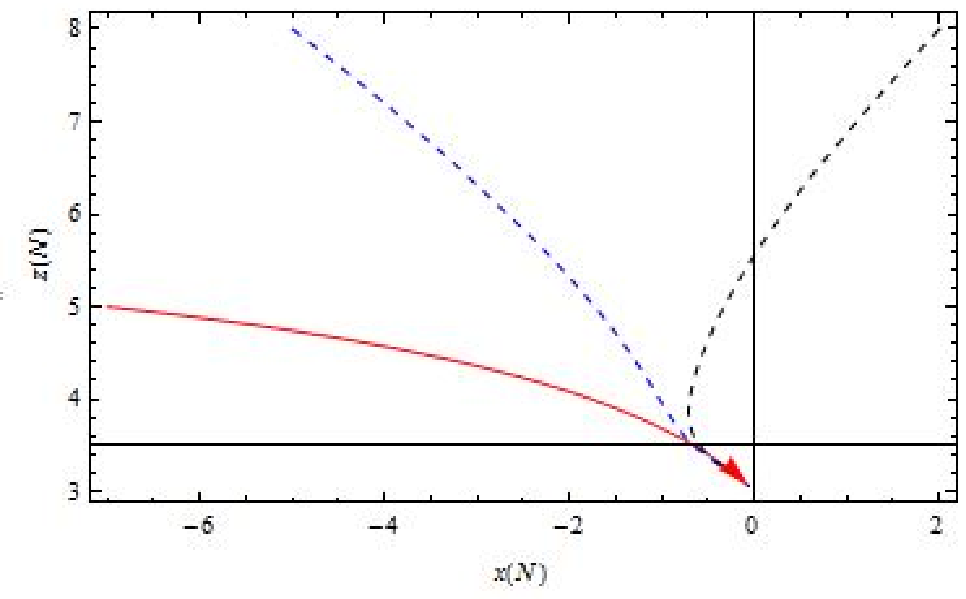}&
		\includegraphics[width=5cm, height=5cm]{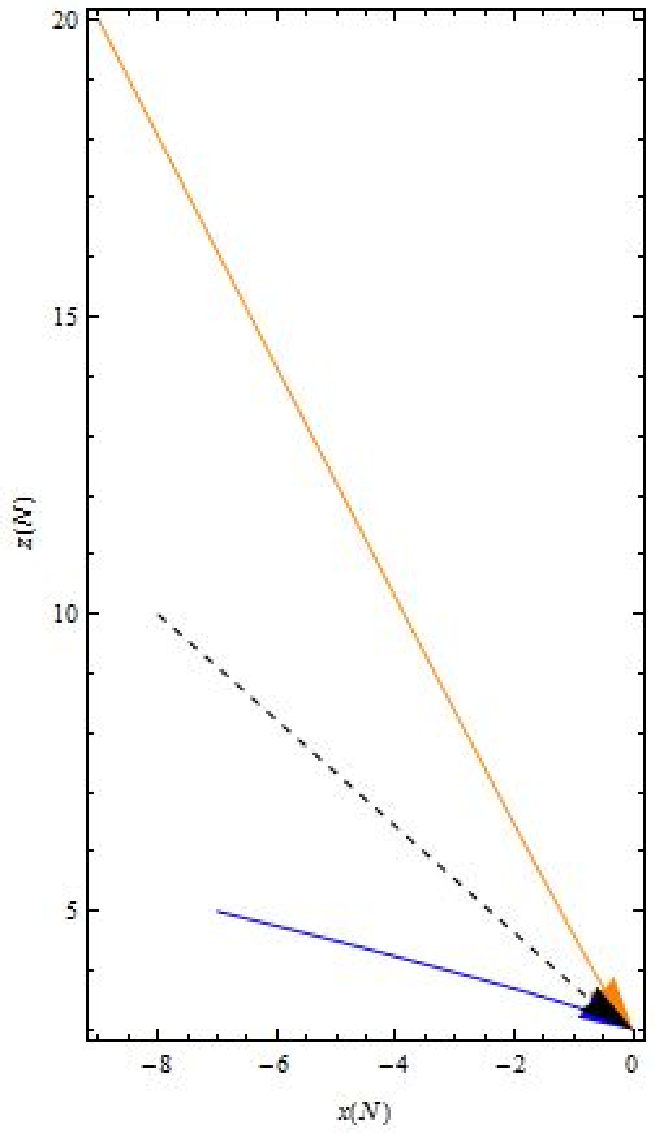}&
		\includegraphics[width=5cm, height=5cm]{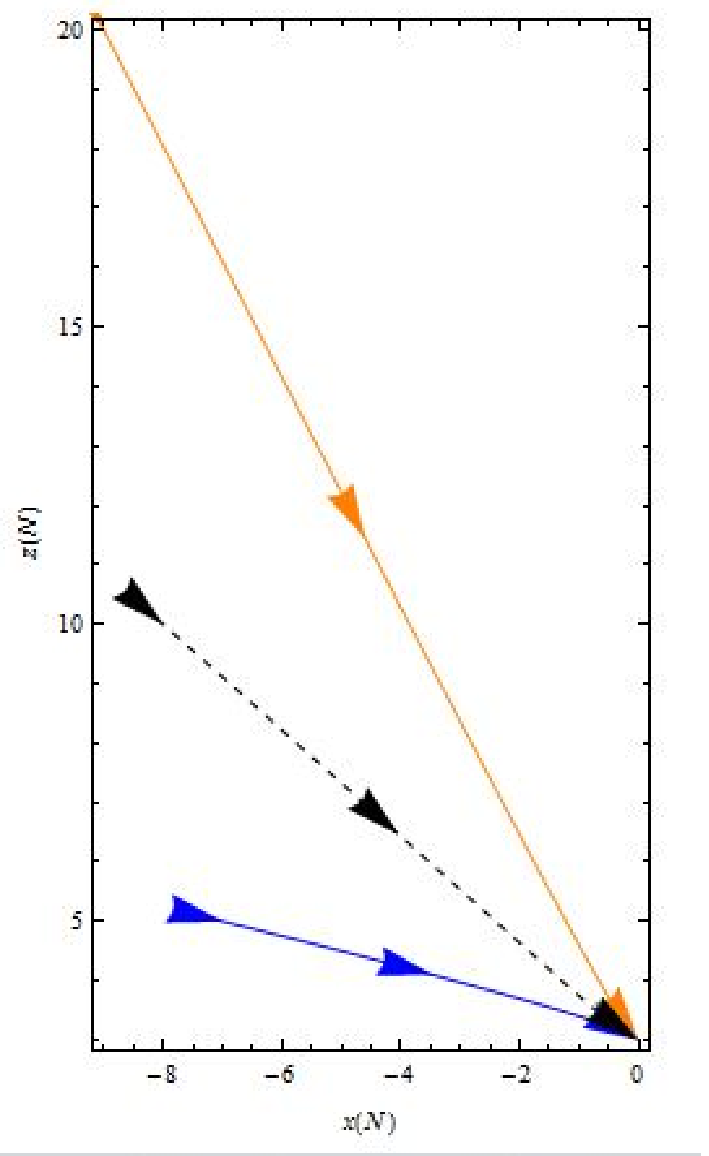}
	\end{tabular}
	\caption{ Parametric and vector plots of the solutions in the plan $x-z$ of dynamical system (\ref{syst1}) for $m=0$}
	\label{fig2}
\end{figure}
\newline
A common conclusion arisen from the analysis of the evolution of the curves presented by the plots of  the figures \ref{fig1} and \ref{fig2} is that the equilibrium $X^3_*= (0 , -3,3)$ is a de Sitter attractor or furthermore, the final de Sitter attractor  of the phase space. 
We  are now looking for illustration of  the global attractive behaviour of this point.  For this, we plot the three dimensional flow for different initial conditions. The obtained figure \ref{fig3} proves the absolute attractor character of the critical point $X^3_*$
\begin{figure}[h]
	\centering
	\begin{tabular}{rl}
		\includegraphics[width=8cm, height=5cm]{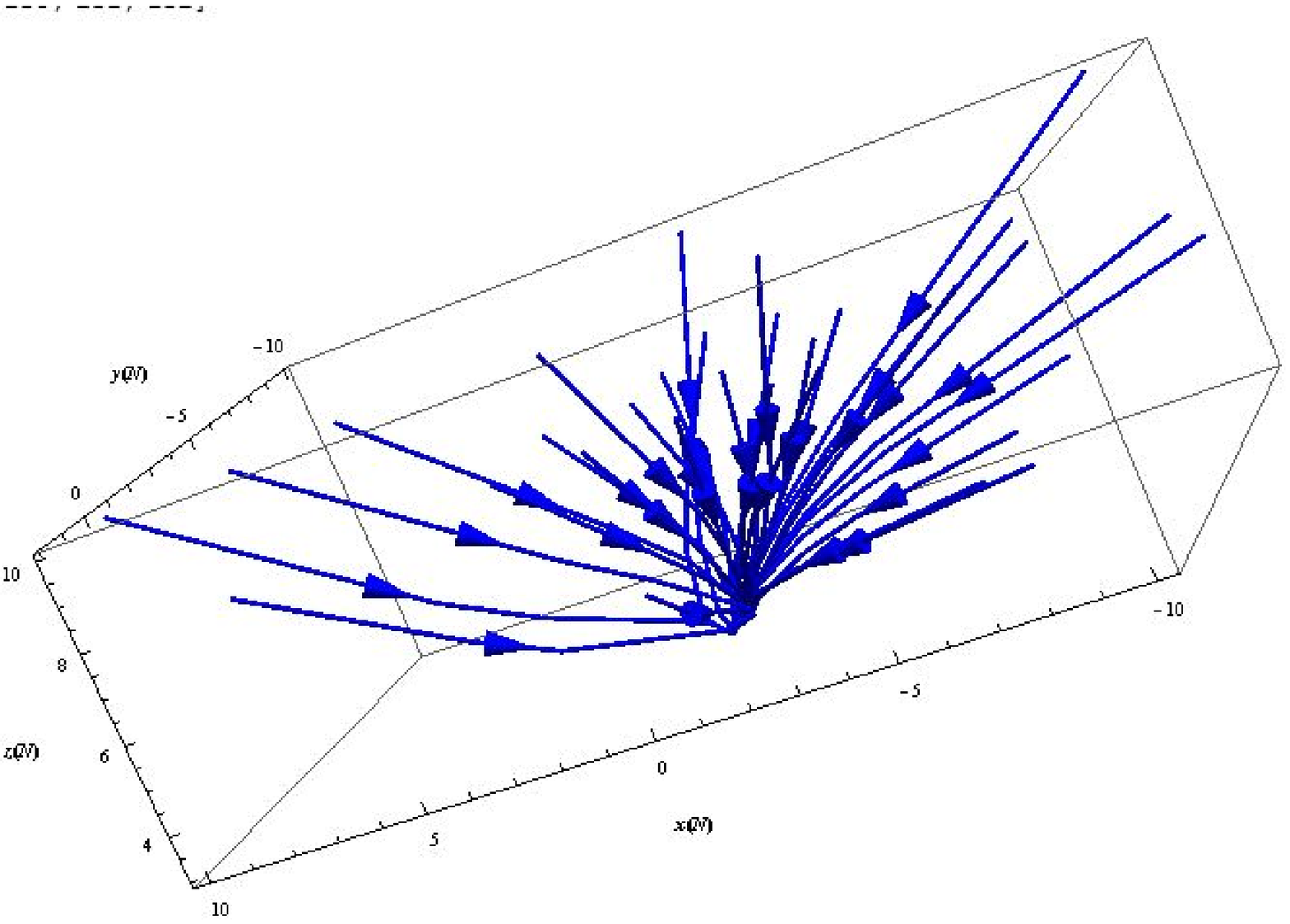}&
		\includegraphics[width=8cm, height=5cm]{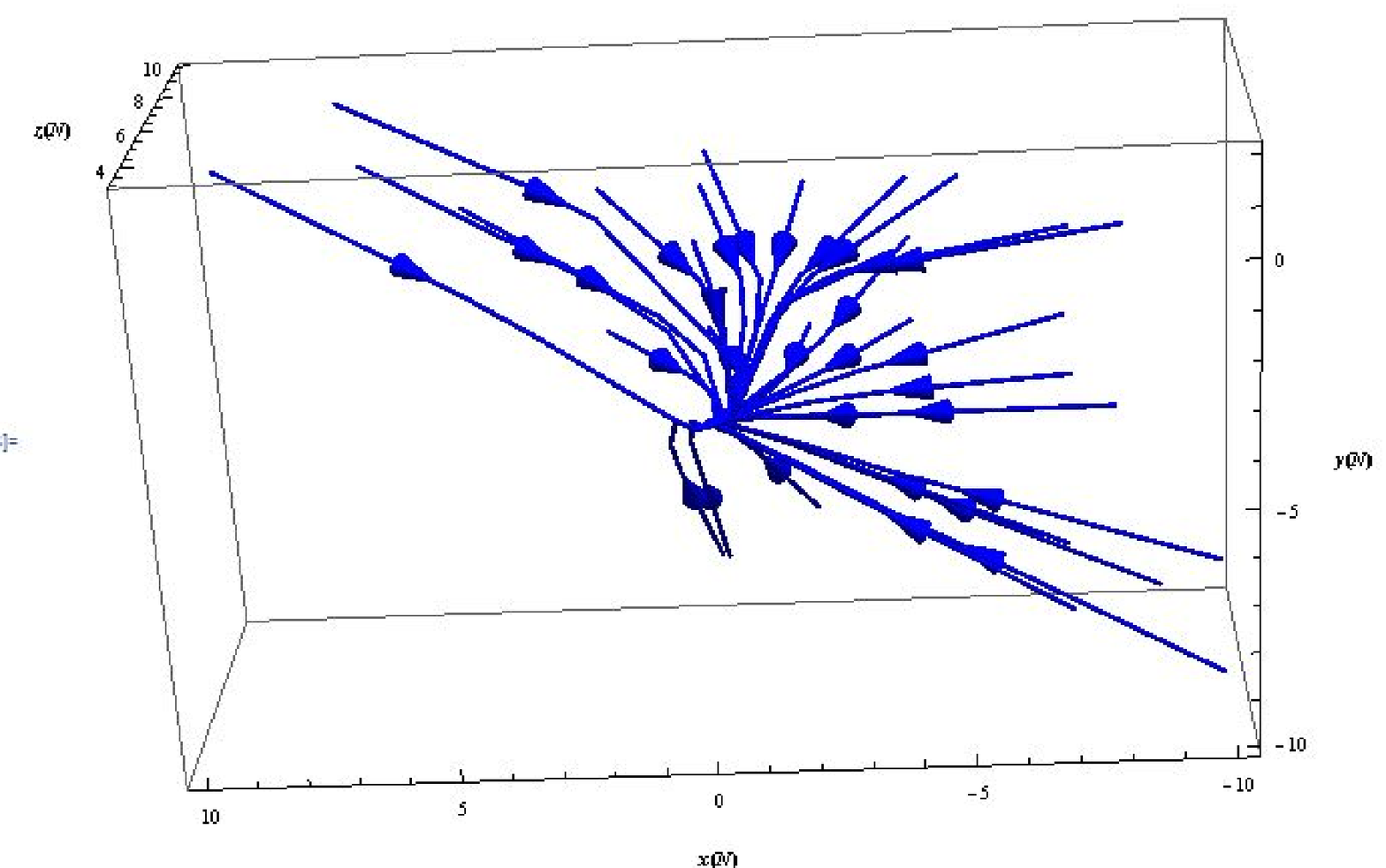}
	\end{tabular}
	\caption{ The three dimensional flow of dynamical system (\ref{syst1}) for different initial conditions and  for $m=0$}
	\label{fig3}
\end{figure}
\newline
 Regarding the dynamic system (\ref{syst1}), the  differential equation in $z$ decouples itself of the other equations of the system. It is also what has facilitated the research of the analytic solutions and justify the order of presentation of these solutions higher.  Indeed, we are going to fix $z$ to its asymptotic value $(z =3)$ and  show the stability  and the attractor fixed point  $X^3_*$. For several initial conditions, we draw the different trajectories  in the plane $x-y$. We note a strong attraction of the different trajectories toward the equilibrium  $X^3_*$  as it was shown by the corresponding figure \ref{fig4}.
\begin{figure}[h]
	\centering
	\includegraphics[width=8cm, height=5cm]{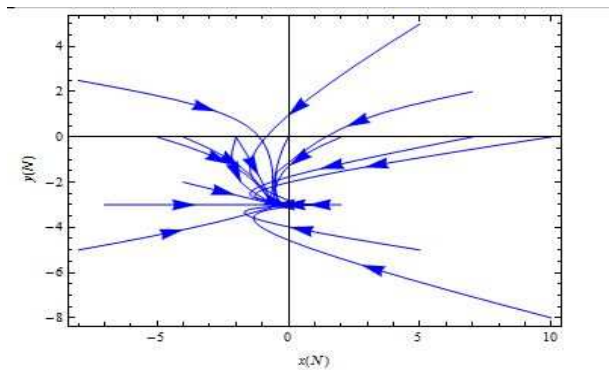}
\caption{ Trajectories in the plane $x-y$ of  dynamical system (\ref{syst1})for $m=0$}
\label{fig4}
\end{figure}
\newline
In summary, our previous results show that for $m=0$, 
the dynamic system (\ref{syst1}) becomes autonomous and admits one de Sitter stable point  $X^3_*=(0 , -3,3)$  (very apparent through the different representations) and two  de Sitter unstable fixed points $X^1_*=(0 , 0,3)$  and $X^2_*=(3 , 0,3)$. These results coincide with those existing in other theories of modified gravity  (for example the interesting work \cite{Sergy}) and align  again the  $f(T)$ theory therefore in the rank of the modified gravitational theories  which can reproduce the de Sitter inflationary era.  
In the framework of our present work, the inflationary evolution
appears by the fact that, in the case  of  quasi-de sitter $(m=0)$, the $f (T)$ theory leads  to a  stable de Sitter attractor which is asymptotically reached when the e-folding number $ N $ stretches toward $60$. We recall here that the previous interpretation  is solely valid in the quasi-de Sitter case namely, the case where the scale factor   takes the form in (\ref{scal}).  Otherwise, in the theory like the symmetric bounce cosmology  \cite{Oikonomou} where the factor of scale is $a(t)=e^{\lambda t^2}$,$\lambda>0$, the parameter $m$ vanishes but the De Sitter attractor in this case can take another physical interpretation: it can be qualified of late-time de Sitter attractor.Presently, we  ask the question to know: what are the cosmological models  $ f (T) $ capable to reproduce the different physical features which result from the previous dynamic system. Follow us in the coming subsection.
       \begin{center}
       	{\bf c- Reconstruction of cosmological vacuum $f(T)$ models}                     	
       \end{center}
Here, we  reconstruct some $f(T)$ models near the previous fixed points which have been obtained in the case $m=0$. We recall also that 
this studied case is the one of which the conditions of slow-roll are verified, so, the reconstructed models must  also take  in account  this aspect. The concerned points are  $X^1_*=(0 , 0,3)$, $X^2_*=(3 , 0,3)$ and $X^3_*=(0 , -3,3)$. \\
Let start with the stable fixed point $X^3_*=(0 , -3,3)$. According to the consideration made in (\ref{pose}), the application of this point leads to the following differential equations
\begin{equation}\label{prem}
 -\frac{\dot{F}(T)}{HF(T)}=0,\qquad \frac{f(T)}{4H^2F(T)}=-3.
 \end{equation}
The first differential equation of (\ref{prem}) breeds
\begin{equation}
\dot{F}(T)=0, \qquad  F(T)\neq 0  \qquad \text{and}  \qquad H(t)\neq 0
\end{equation}
Since, we are performing in the case where $m=0$, which makes  true the slow-roll conditions $\ref{slow}$, we shall have $H\neq 0$, and one gets
\begin{equation}\label{fg}
\dot{F}(T)=0\Longrightarrow-H_i\frac{d^2f}{d^2T}=0 \Longrightarrow f(T)=\lambda_1T+\lambda_2,
\end{equation}
where $\lambda_1$ and  $\lambda_2$ are constants of integration. By setting $ \lambda_1=1$ and  $\lambda_2=\Lambda$, with  $\Lambda$, the well 
known cosmological constant \cite{Nottale}, one obtains the following model
\begin{equation}\label{god1}
f(T)=T+\Lambda,
\end{equation} 
which corresponds to the density Lagrangian of Tele-Parallel coupled with the constant cosmology. A de Sitter universe is a cosmological solution to the Einstein field equations of general relativity or Tele-Parallel theory, named after Willem de Sitter. It models the universe as spatially flat and neglects ordinary matter $(m=0)$, so the dynamics of the universe are dominated by the cosmological constant, thought to correspond to dark energy in our Universe or the inflaton field in the early Universe. So, these cosmological features can be reproduced by the model in (\ref{god1}).\par  
The general $f(T)$ forms obtained in  (\ref{fg}), satisfied the second differential equation of  (\ref{prem}) via the following relation 
\begin{equation}\label{con}
\frac{\lambda_2}{-6\lambda_1H^2}\simeq 0.
\end{equation}
By the fact that $\lambda_1$ and $\lambda_2$ integration constants, the condition (\ref{con}), which  required large valor of hubble parameter, holds true when the slow-roll conditions are satisfied. Consequently the  $f(T)$ gravity solution of (\ref{fg}) can generates
 the quasi-de Sitter evolution  that yields $m=0$ in the large scalar torsion era because of the required large hubble parameter. As conclusion,  at the large scalar torsion era, the $f(T)$ form in  (\ref{fg})  can lead  to the de Sitter stable point  $X^3_*=(0 , -3,3)$. \par
 Otherwise,  by an inverse analysis to the previous one, let's consider the second differential equation of  (\ref{prem}). Its resolution yields
 \begin{equation} \label{imp}
 	\frac{f(T)}{4H^2F(T)}=-3\Longrightarrow f(T)=2T\frac{df(T)}{dT} \Longrightarrow f(T)=\sqrt{-T}+ \delta,
 \end{equation}
%%%%%%%%%%%%%%%%%%%%%%%%%%%%%%%%%%%%%%%%%%%%%%%%%%%%%%%%%%%%%%%%%%%%% 
 where $\delta$ is an integration constant.  Let's note that the generated  square root model in (\ref{imp}) has a strong cosmological implication in general and in the survey of cosmological dynamical systems. This $f(T)$ model can be
 recovered via reconstruction scheme of holographic dark
 energy \cite{Daouda}. Also it can be inspired from a model for
 dark energy model form  of the Veneziano ghost \cite{Karami}. Furthermore, and according to \cite{Jamil}, the first term  (the square root one) of this model denotes a ghost dark energy and
 performs a role of cosmological constant. It can also be reconstructed mathematically as a
 toy model of a type of ghost dark energy if and only if we neglect all matter fields. Recently, attractor solutions for the dynamical system with three fluids (dark matter, dark energy and radiation) interacting non-gravitationally have been investigated to resolve the coincidence problem using similar $f(T)$ \cite{Momeni}.  This reconstructed $f(T)$ model was also  especially at  the heart of  interesting investigation like \cite{Mubasher} where it leads to an
 attractor solution to the dynamical $f(T)$ Friedmann equations of the interacting dark energy model in $f(T)$ cosmology \par 
 The $f(T)$ form in  (\ref{imp}), introduced in the first equation of  (\ref{prem}), leads to 
\begin{equation}
\frac{1}{H^3}\simeq 0,
\end{equation}
which can be met in the large scalar torsion eras. One  again, the $f(T)$ solution in (\ref{imp}), approaches the de Sitter attractor $X^3_*=(0 ,-3,3)$ in the large scalar torsion era.\par  
 By taking into consideration its two first coordinates,  the second fixed point  $X^2_*=(3 ,0,3)$, yields the following system of differential equations 
\begin{equation}\label{deux}
-\frac{\dot{F}(T)}{HF(T)}=3,\qquad \frac{f(T)}{4H^2F(T)}=0,
\end{equation}
where the first one gives
\begin{equation}\label{repas}
 -\frac{\dot{F}(T)}{HF(T)}=3 \Longrightarrow H_i\frac{d^2f(T)}{d^2T}-\frac{df(T)}{dT}=0 \Longrightarrow f(T)=4H_ie^{\frac{T}{4H_i}}\sigma_1+\sigma_2. 
\end{equation}
$\sigma_1$ et $\sigma_2$ are integration constants. Such a model in (\ref{repas}) is, according to the literature, the so-called exponential $f(T)$ model which, for example,  can tend to the $\Lambda CDM$ cosmology \cite{Nesseris} or produces an inflationary scene in absence of all matter fields \cite {El}. In large scalar torsion, the model   in (\ref{repas}) can verify the second differential equation   in (\ref{deux}) if the slow-roll condition holds true. Indeed,  one has 
 \begin{equation}
\frac{f(T)}{4H^2F(T)}=\frac{4H_ie^{\frac{T}{4H_i}}\sigma_1+\sigma_2 }{4H^2\sigma_1 e^{\frac{T}{4H_i}}}\sim \frac{H_i}{H^2}\simeq 0
\end{equation}
As conclusion, the exponential $f(T)$ solution in (\ref{repas}), can reproduce the de Sitter unstable point  $X^2_*=(0 ,-3,3)$ in the large scalar torsion era.\par 
In the case of the first unstable fixed point  $X^1_*=(0 ,0,3)$ , the corresponding system of differential equations is
\begin{equation}\label{trois}
-\frac{\dot{F}(T)}{HF(T)}=0,\qquad \frac{f(T)}{4H^2F(T)}=0.
\end{equation}
After some calculations, one can conclude that the model in (\ref{fg}) can fit this fixed point  $X^1_*=(0 ,0,3)$ in the same conditions as the previous one. \par 
In another theory of gravity, like the well-known $f(R)$ gravity, this cosmological studying of autonomous dynamical system in the quasi-de Sitter inflation yields in addition to exponential $f(R)$ form, a very interesting inflationary model which is the $R^2$ gravity 
\cite{Sergy}, with $R$ the scalar curvature. They show that the presence of an $R^2$ term in $f(R)$ gravity leads
to instability, which may be viewed as an indication of the graceful exit from the inflationary era. Otherwise, the authors in \cite{Bamba} have demonstrated that the de Sitter inflation with graceful exit can be realized in $T^2$ while  $R^2$ may lead to quasi-de Sitter inflation with graceful exist.  As we can already note it in the present work where the fixed points resulted from a quasi-de Sitter's approach, any mathematical approximation did not lead to the model $ T^2 $. The question that we are  asking ourself, is to know if a $T^2$ model can reproduce the behavior of the obtained dynamical. 

\begin{figure}[h]
 	\centering
 	\includegraphics[width=8cm, height=5cm]{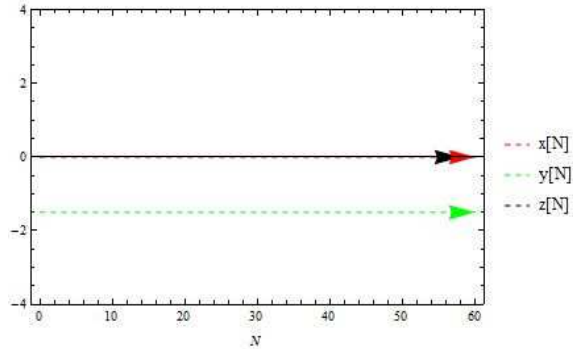}
 	\caption{Dynamical evolution of the variables $(x(N),y()N),z(N)$ in $T^2$ gravity for $H_i=10^{20}\text{sec}^1$ and  $H_0=10^{13}\text{sec}^2$ \cite{Odintsov}} 
 	\label{fig5}
 \end{figure}

                 \begin{center}
                 	{\bf d- Testing an inflationary graceful exit  $f(T)$ model: the $T^2$ model}                     	
                 \end{center}
 Let's consider the following functional general form of the $T^2$ gravity 
 \begin{equation}\label{ora}
 	f(T)=T+\frac{1}{\alpha H_i}T^2,
 \end{equation}
 where the parameter $H_i$  has dimensions of $\text{mass}^2$ and $\alpha$ is dimensionless parameter \cite{Sergy}- \cite{Bamba}.
This $f(T)$ model is called to satisfy  the vacuum FRW equations and lead to the Hubble parameter expression presented in (\ref{scal}) which corresponds to $m=0$ or to slow-roll conditions. Accordingly, We are searching in the first time the expression of the parameter $\alpha$  in order the make filling the previous conditions,  the  $T^2$ gravity. Inserting the $f(T)$ function in  the second 
 vacuum Friedmann equation in (\ref{e4}), one obtains 
  \begin{equation}\label{ore}
  \frac{1}{2}T+\frac{1}{2\alpha H_i}T^2+ (6H^2+2\dot{H})(1+\frac{2}{\alpha H_i}T)-\frac{48}{\alpha H_i}H^2\dot{H}=0
 \end{equation}
 By making using the slow-roll condition ($\dot{H}\ll H^2$), and under the initial condition $H[t_k]=H_0$ (approach of \cite{Sergy}), we solve
  this differential equation and the approximate Hubble parameter,  solution of this equation reads 
  \begin{equation}
  	H(t)\backsimeq H_0+\frac{1}{16}(-18H_0^2+\alpha H_i)(t-t_k).
  \end{equation}
 In  order to realize the quasi-de sitter evolution, the Hubble parameter must be recast in the form \cite{Sergy}
 \begin{equation}
 	H(t)\backsimeq H_0-H_i(t-t_k).
 \end{equation}
So
  \begin{equation}
  \alpha=-16+18\frac{H_0^2}{H_i}.
 \end{equation}
 From \cite{Odintsov}, the dimensions of $H_0$ and $H_i$ maintain the nature of $\alpha$ (dimensionless parameter). The corresponding functional
  of $T^2$ gravity is 
\begin{equation}\label{fin}
f(T)=T+\frac{1}{(-16+18\frac{H_0^2}{H_i}) H_i}T^2.
\end{equation}
This cosmological model, which is sensible of realizing the quasi-de Sitter evolution, can be applied  to  the dynamical  variables ($x,y,z$), making them, the explicit function of e-folding number. But to arrive there, we must also try to express the cosmic time  as function of e-folding number; $t=(H_0+\sqrt{H_0^2-4NH_i})/(2H_i)$ and then, we obtain
\begin{eqnarray}
x(N)=\frac{12 H_i}{(-8+6 n) H_i+3 H_o \left(2 H_o+\sqrt{-4 n H_i+H_o^2}\right)},\\
y(N)=\frac{6 (8-3 n) H_i-9 H_o \left(5 H_o+\sqrt{-4 n H_i+H_o^2}\right)}{8 (-4+3 n) H_i+12 H_o \left(2 H_o+\sqrt{-4
		nH_i+H_o^2}\right)},\\
z(N)=\frac{2 n(1+3 n) H_i-H_o \left(H_o+\sqrt{-4 n H_i+H_o^2}\right)}{2 n^2 H_i}	
\end{eqnarray}   
 In figure \ref{fig5}, we plot the evolutions of these variables 
 \newline
 Any fixed point among the three  fixed points for $m=0$ is not reached in the dynamical evolution described in the figure \ref{fig5}. The convergence is made in the direction of a new point $(0,-3,0)$. Let remark here that, the two first coordinates of this point which correspond to those of the de sitter stable point  $X^3_*=(0 ,-3,3)$, have led to viable de sitter inflationary vacuum models (\ref{fg})-(\ref{imp}). But the contradiction here comes from the third component ($z=0$) giving $\omega_{eff}=1$ which has nothing with de Sitter evolution. The quadratic $f(T)$ model can not reproduce the quasi-de Sitter evolution contrarily to $f(R)$ gravity as it was investigated in \cite{Bamba}.   
 
   \begin{center}
   	{\bf e-Studying matter dominated in vacuum $f(T)$: $m=-9/2$}                     	
   \end{center}
 Secondly to the case $m=0$, we consider $m=-9/2$  which corresponds to the matter dominated Universe. The corresponding Jacobian matrix reads 
  \begin{eqnarray}
 \mathcal{J}=\Bigg[\begin{array}{ccc}
 3 - y - 2 z + 2 x & 3 - x + 2 y & 6 - 2 x - 2 z \\ 
 0 & -z + 6 + 2 y & -3 - y \\ 
 0 & 0 & 12 - 4 z
 \end{array} \Bigg],
 \end{eqnarray}
 with the functions $g_i$ given by  
 \begin{eqnarray*}
 	g_1&=&  - 9/2 + 6 z + 3 y + 3 x - 2 zx - yx - z^2 + x^2 + y^2\\
 	g_2&=& 9-3z-yz+6y+y^2\nonumber\\
 	g_3&=& -27/2+12z-2z^2 \nonumber
 \end{eqnarray*}
 The fixed points and the matrix eingenvalues for each of them are presented  in the following table 
 \begin{center}
 	\begin{tabular}{|c|c|c|c|c|c|c|}
 		\hline Fixed points & $x$  & $y$ & $z$ & $\eta_1$ &  $\eta_2$ &  $\eta_3$  \\ 
 		\hline $X^1_*$ & $-\frac{3}{2}$ & $-\frac{3}{2}$ & $\frac{3}{2}$ & $6$ & $-\frac{3}{2}$ & $\frac{3}{2}$ \\ 
 		\hline $X^2_*$  & $0$ & $-\frac{3}{2}$ & $\frac{3}{2}$ & $6$ & $\frac{3}{2}$ & $\frac{3}{2}$ \\ 
 		\hline $X^3_*$ & $-\frac{3}{2}$ & $-3$ & $\frac{3}{2}$ & $6$ & $0$ & $-\frac{3}{2}$ \\ 
 		\hline $X^4_*$ & $\frac{3}{2}$ & $\frac{3}{2}$ & $\frac{9}{2}$ & $-6$ & $-\frac{9}{2}$ & $\frac{9}{2}$ \\ 
 		\hline $X^5_*$ & $6$ & $\frac{3}{2}$ & $\frac{9}{2}$ & $-6$ & $\frac{9}{2}$ & $\frac{9}{2}$ \\ 
 		\hline $X^6_*$ & $\frac{3}{2}$ & $-3$ & $\frac{9}{2}$ & $-6$ & $0$ & $-\frac{9}{2}$ \\ 
 		\hline 
 	\end{tabular} 
 \end{center}
 \vspace{1cm}
 All the equilibria presented in this table are non-hyperbolic and only the last is seemed to be stable. We can also remark that for the set of these fixed points, two values of the third coordinates can be distinguished: $z=3/2$ and $z=9/2$. By inserting the first value, $z=3/2$, in 
 (\ref{ome}), one obtains the effective EoS parameter which reads $\omega_{eff}=0$. This value holds physically and corresponds really to matter dominated evolution. As for the second value,  $z=9/2$, it leads to $\omega_{eff}=-2$ which  has no physical significance, since
 it corresponds to a phantom evolution. Furthermore, this value is asymptotic value of the dynamical system for $m=-9/2$. The reason is that, the analytical solution of the $z(N)$ differential equation of the dynamical system for $m=-9/2$ reads 
 \begin{equation}\label{corec}
 z(N)=\frac{3(3e^{6N}-\tau)}{2(e^{6N}-\tau)},  
 \end{equation}\label{pg}
where $\tau$ stays for an integration constant. For large value of e-folding number, $z(N)\longrightarrow 9/2$. Using (\ref{corec}), one can also obtain the analytical expression of $y(N)$ excepting those of $x(N)$ whose differential equation is not easy to solve for $m=-9/2$.
\begin{eqnarray}
 y(N)&=& -\Bigg[ \frac{3}{\tau^{1/2}} \sqrt{e^{6 N}}+
\phi \Bigg(\frac{3}{2} (\frac{-1}{\tau})^{1/4} \left(e^{6 N}\right)^{1/4}  \text{Hypergeometric2F1}\left[-\frac{1}{4},\frac{1}{2},\frac{3}{4},\frac{1}{\tau}e^{6N}\right]+\nonumber\\
&&+\frac{3}{2} (\frac{-1}{\tau})^{1/4} \left(e^{6 N}\right)^{1/4} \left(-\frac{1}{\sqrt{1-\frac{e^{6 N}}{\tau}}}+\text{Hypergeometric2F1}\left[-\frac{1}{4},\frac{1}{2},\frac{3}{4},\frac{1}{\tau}e^{6N}\right]\right)\Bigg)\Bigg]\nonumber\\
&&\times \Bigg[ (\frac{1}{\tau})^{1/2} \sqrt{e^{6 N}}+ \phi(\frac{-1}{\tau})^{1/4} \left(e^{6 N}\right)^{1/4}
\text{Hypergeometric2F1}\left[-\frac{1}{4},\frac{1}{2},\frac{3}{4},\frac{1}{\tau}e^{6N}\right]\Bigg]^{-1},
\end{eqnarray}
 with $\phi$ an integration constant. Despite the fact that $x(N)$ is not analytically, one can find out the behaviours of these dynamical variables by proceeding to an numerical solving. \\
  For two different initials conditions, we plot the behaviour of these solutions in the following diagrams of figure \ref{fig6}:
% \begin{figure}[h]
 %	\centering
 %	\begin{tabular}{rl}
 %		\includegraphics[width=8cm, height=5cm]{cap11.eps}&
 %		\includegraphics[width=8cm, height=5cm]{cap12.eps}
 %	\end{tabular}
 %	\caption{Behaviour of of dynamical system (\ref{syst1}) solutions for $m=-9/2$  and for two initial conditions: $x(0)=-7$, $y(0)=0$, $z(0)=5$, (left plot) and  $x(0)=0$, $y(0)=0$, $z(0)=0$ (right plot)} .
 %	\label{fig6}
 % \end{figure}
 \newline
These plots show that in the case where the initial condition is  $x(0)=-7$, $y(0)=0$, $z(0)=5$, the stable fixed point
 $X^6_*=( \frac{3}{2}, -3,\frac{9}{2})$  is asymptotically reached where as  in the second case, $x(0)=0$, $y(0)=0$, $z(0)=0$, the dynamical system solutions simply blow-up because the  variable $z$ deviates from  its equilibrium value (see the curve in black in the two plots).
 By fixing $z=9/2$, we plot  in the figure 
 %\ref{fig7},
  the trajectories  in the plane $x$-$y$ for various initial conditions.  
 %\begin{figure}[h]
 %	\centering
 %	\includegraphics[width=8cm, height=5cm]{cap13.eps}
 %	\caption{ Trajectories in the plane $x-y$ of dynamical system (\ref{syst1}) for various initial conditions and  for $m=-9/2$}
 %	\label{fig7}
 % end{figure}
 \vspace{1cm}
 The correlation among the three solutions can be viewed through the following three dimensional plot in \ref{fig8} 
% \begin{figure}[h]
 %	\centering
 %	\begin{tabular}{rl}
 %		\includegraphics[width=8cm, height=5cm]{cap14.eps}&
 %		\includegraphics[width=8cm, height=5cm]{cap15.eps}
 %	\end{tabular}
 %	\caption{ The three dimensional flow of dynamical system (\ref{syst1}) for various initial conditions  and for $m=-9/2$}
 %	\label{fig8}
 % \end{figure}
  \vspace{1cm}
 As conclusion, the autonomous dynamical corresponded to $m=-9/2$ leads  to six non-hyperbolic fixed points. Three of these points, $X^1_*=( -\frac{3}{2}, -\frac{3}{2},\frac{3}{2})$, $X^2_*=( 0, -\frac{3}{2},\frac{3}{2})$ and   $X^3_*=( -\frac{3}{2}, -3,\frac{3}{2})$, are unstable but describe really the matter dominated era because they are characterized by the $z=\frac{3}{2}$ which yields $\omega_{eff}=0$.  
 Nevertheless, the three others fixed points among which, two are unstable:$X^4_*=( \frac{3}{2}, \frac{3}{2},\frac{9}{2})$, $X^5_*=( 6, \frac{3}{2},\frac{9}{2})$, and the last one is attractor:$X^6_*=( -\frac{3}{2}, -3,\frac{9}{2})$,  are distinguished by  $z=\frac{9}{2}$
 and so describe the phantom evolution because for this value of $z$ the effective EoS parameter becomes $\omega_{eff}=-2$. With this aspect, the $f(T)$ theory can  be useful for a dark energy description because from an inflationary point of view, it is not so appealing having a phantom theory at hand. Unfortunately, the only stable fixed  point, as shown by the figures 
 %\ref{fig6} and \ref{fig7}
  is a phantom point 
 $X^6_*=( -\frac{3}{2}, -3,\frac{9}{2})$. It is not too astonishing because, $y=-3$ yields a $f(T)$ model in (\ref{imp}), very literally  known 
 as a ghost dark energy model. However, the figure 
 %\ref{fig8},
  which presents the three dimensional flow for various initials conditions, shows  that the attractor equilibrium  $X^6_*=( -\frac{3}{2}, -3,\frac{9}{2})$ can be reached or not. Indeed, regarding this figure, it exists one flow which evades this point. This fact reassures us that if we add more initial conditions, one will have luck to reach other unstable $f(T)$ points leading to the matter dominated era. So, in vacuum $f(T)$ description, the instability leads the matter dominated era. We will end phase space analysis on the case where a perfect fluid like ordinary matter or radiation are coupled to $f(T)$ gravity known as the non-vacuum $f(T)$ gravity.  
 
 \section{  Autonomous Dynamical System in Non-vacuum $f(T)$ Gravity}
 The non-vacuum  $f(T)$ gravity, in the space-time of Friedmann-Roberson-Walker is governed by the motion equations presented in (\ref{e1}) and
 (\ref{e2}). The energy density $\rho$ and the pressure  $P$ present in these equations will characterise a perfect fluid composed of ordinary matter and radiation. Like the previous section, our autonomous dynamical system approach is based on the second Friedmann  equations. Since the pressure of ordinary matter is zero, so only radiation pressure ($P_r$) will be taken into account. For this reason, we pose the following  four  dynamical variables: 
 \begin{eqnarray}\label{pose2}
 x=-\frac{\dot{F}(T)}{HF(T)},\qquad y=\frac{f(T)}{4H^2F(T)}, \qquad z=\frac{\Re}{H^2}, \qquad u=-\frac{\kappa^2P_r}{2H^2F(T)}.
 \end{eqnarray}
 The corresponding dynamical system reads
 \begin{eqnarray}\label{syst2}
 \frac{dx}{dN}&=& -m - 9 + 6 z + 3 y + 3 x - 2 zx - yx + yu - z^2 + x^2 + y^2\\
 \frac{dy}{dN}&=& 9 - 3 z - yz + 6 y + yu + y^2\nonumber\\
 \frac{dz}{dN}&=& -m-18+12z-2z^2 \nonumber\\
 \frac{du}{dN}&=& ux - 2 zu + 6 u\nonumber
 \end{eqnarray} 
 with $m$ the same parameter as defined in the previous section. For any value of the parameter $m$ the fixed points are presented in the following table \\
 \begin{center}
  \begin{tabular}{|c|c|c|c|c|c|c|}
  	\hline Critical points  & $x$ & $y$ & $z$&$u$ \\ 
  	\hline $X^1_*$ &$ -\frac{\sqrt{-m}}{\sqrt{2}}$ & $-3$&  $3-\frac{\sqrt{-m}}{\sqrt{2}}$& $0$\\
  	\hline $X^2_*$ &$\frac{\sqrt{-m}}{\sqrt{2}}$&$-3$& $3+\frac{\sqrt{-m}}{\sqrt{2}}$& $0$\\
  	\hline$X^3_*$ & $-\sqrt{2} \sqrt{-m}$& $-3-\frac{\sqrt{-m}}{\sqrt{2}}$& $3-\frac{\sqrt{-m}}{\sqrt{2}}$&$ \frac{3 \left(3 \sqrt{2} \sqrt{-m}+m\right)}{18+m}$ \\ 
  	\hline$X^4_*$  &$\sqrt{2} \sqrt{-m}$& $-3+\frac{\sqrt{-m}}{\sqrt{2}}$& $3+\frac{\sqrt{-m}}{\sqrt{2}}$& $\frac{3 m}{3 \sqrt{2} \sqrt{-m}+m}$ \\ 
  	\hline $X^5_*$  & $3-i \sqrt{2} \sqrt{m}$& $-\frac{i \sqrt{m}}{\sqrt{2}}$& $3-\frac{i \sqrt{m}}{\sqrt{2}}$& $0$\\ 
  	\hline$X^6_*$ &$ -\frac{i \sqrt{m}}{\sqrt{2}}$& $-\frac{i \sqrt{m}}{\sqrt{2}}$& $3-\frac{i \sqrt{m}}{\sqrt{2}}$&$ 0$\\ 
  	\hline $X^7_*$ &$ 3+i \sqrt{2} \sqrt{m}$& $\frac{i \sqrt{m}}{\sqrt{2}}$&$ 3+\frac{i \sqrt{m}}{\sqrt{2}}$&$ 0$\\
  	\hline $X^8_*$ & $\frac{i \sqrt{m}}{\sqrt{2}}$& $\frac{i \sqrt{m}}{\sqrt{2}}$& $3+\frac{i \sqrt{m}}{\sqrt{2}}$&$ 0$\\
  	\hline
  \end{tabular}
 \end{center}
The Jacobian matrix which linearises the system (\ref{syst2}) doesn't depend from the parameter $m$ and its reads 
 \begin{eqnarray}
 \mathcal{J}=\Bigg[\begin{array}{cccc}
3 - y - 2 z + 2 x& 3 - x + 2 y + u& 6 - 2 x - 2 z& y \\ 
0& -z + 6 + 2 y + u& -3 - y& y \\ 
 0& 0& 12 - 4 z& 0\\
 u& 0& -2 u& x + 6 - 2 z
 \end{array} \Bigg]
 \end{eqnarray}
 It is obtained from the  following    general $g_i$ functions
 \begin{eqnarray}
 g_1&=& -m - 9 + 6 z + 3 y + 3 x - 2 zx - yx + yu - z^2 + x^2 + y^2\\
  g_2&=& 9 - 3 z - yz + 6 y + yu + y^2\nonumber\\
 g_3&=& -m-18+12z-2z^2 \nonumber\\
  g_4&=& ux - 2 zu + 6 u\nonumber
 \end{eqnarray} 
 Different values of parameter $m$ will lead to various features of the dynamical system (\ref{syst2}) which reveals autonomous when $m$ is constant. We start the analysis of the system with $m=0.$
     \subsection{de Sitter Inflationary Attractors and their Stability in non-vacuum $f(T)$ gravity: $m\backsimeq0$}
 For $m\backsimeq0$, we are going to analyse the behaviour of the space phase in non-vacuum $f(T)$ gravity. The corresponding fixed points and the associate eigenvalue of Jacobian matrix are given in the following table\\
\begin{center}
 \begin{tabular}{|c|c|c|c|c|c|c|c|c|}
 	\hline Fixed points & $x$  & $y$ & $z$ &$u$& $\eta_1$ &  $\eta_2$ &  $\eta_3$ &  $\eta_4$ \\ 
 	\hline $X^1_*$ & $0$ & $-3$ & $3$& $0$ & $0$ & $-3$ & $ 0$& $0$ \\ 
 	\hline $X^2_*$  & $3$ & $0$ & $3$ &$0$& $0$ & $3$ & $3$&$3$ \\ 
 	\hline $X^3_*$ & $0$ & $0$ & $3$ &$0$& $0$ & $-3$ & $0$& $3$ \\ 
 	\hline 
 \end{tabular} 
\end{center}
 In this previous table are set thee non-hyperbolic fixed points which can be classified in two categories according to their associated eigenvalue: two unstable points $X^2_*=(3, 0,3,0)$ and $X^3_*=(0, 0,3,0)$ and one stable point $X^1_*=(0, -3,3,0)$. All of them are characterised by  $z=3$ which implies $\omega_{eff}=-1$. Consequently all these fixed points are the de Sitter equilibria. For $m=0$, only the third differential equation of (\ref{syst2}) can be analytically solved  and the solution is the same as those presented in  (\ref{caf}). In parallel way as the previous section, we are going to analyse numerically the solution behaviours of (\ref{syst2}) and to find out their possible  physical significance. Firstly, by considering the initial condition $x(0)=-7$, $y(0)=-3$, $z(0)=5$,$u(0)=-1$, we plot in figure 
 %\ref{fig9},
   the evolution of  these solutions for two choices of e-folding number: $N\in [0,2]$ (left plot) and $N\in [0,60]$ (right plot).The evolutions presented in  the two plots of figure %\ref{fig9},
    show  that the de Sitter attractor  $X^1_*=(0, -3,3,0)$ is not directly but asymptotically reached. 
 
 % \begin{figure}[h]
  %	\centering
  %	\begin{tabular}{rl}
  %		\includegraphics[width=8cm, height=5cm]{cap16.eps}&
  %		\includegraphics[width=8cm, height=5cm]{cap17.eps}
  %	\end{tabular}
  %	\caption{ The figure shows the behaviour of solutions of the dynamical system \ref{syst2} for initial conditions $x(0)=-7$, $y(0)=-3$, $z(0)=5$,$u(0)=-1$    and for $m=0$}
  %	\label{fig9}
  % \end{figure}
  \vspace{1cm}
  The behaviours of the dynamical system \ref{syst2} for $m=0$ confirm partially the stability of the fixed point $X^1_*=(0, -3,3,0)$, which means that regardless the initial conditions used, the trajectories are attracted to this de Sitter equilibrium. We shall show more  this interpretation by proceeding to the dimensional plot. We emphasis here that by posing $u(N)=0$, the system \ref{syst2} is restricted to the 
  vacuum $f(T)$ system case which has already been performed in the previous section. For this reason and  in order to show deeply the stability of  the  fixed point  $X^1_*=(0, -3,3,0)$, we draw in figure 
  %\ref{fig10}
   the trajectories of the dynamical system  for various initial conditions in the following three planes: 
  $x-u$ plane (left plot), $y-u$ plane (right plot) and $z-u$ plane (bottom plot). 
  %\begin{figure}[h]
  %	\centering
  %	\begin{tabular}{rll}
  %		\includegraphics[width=8cm, height=5cm]{cap19.eps}&
  %		\includegraphics[width=8cm, height=5cm]{cap18.eps}\\
  %	\includegraphics[width=8cm, height=5cm]{cap20.eps}
  %	\end{tabular}
  %	\caption{ Trajectories of the dynamical system (\ref{syst2}) for $m=0$ in three different planes and for   various initial conditions}
  %	\label{fig10}
  %\end{figure}
 The three plots show clearly that in each plane, the restricted coordinates of the equilibrium $X^1_*=(0, -3,3,0)$  are reached. The de Sitter attractor property of this  fixed point is once again found out. As it was higher performed, we are showing now the global de Sitter attractive of this point through the three dimensional. Among the four dynamical parameters, only $z$ has easily an analytical expression which shows that for large value of e-folding number, $z$ yields $3$. So, we fix $z=3$ and plot via the figure 
 %\ref{fig11},
  the three dimensional flow of   dynamical system (\ref{syst2}) for $m=0$ and for various initial conditions. The plot shows that the equilibrium is a stable de Sitter attractor.\\
  %\begin{figure}[h]
  %	\centering
  %	\begin{tabular}{rll}
  %		\includegraphics[width=8cm, height=5cm]{cap21.eps}&
  %		\includegraphics[width=8cm, height=5cm]{cap22.eps}
  %		  	\end{tabular}
  %	\caption{ Three dimensional flow of   dynamical system (\ref{syst2}) for $m=0$ and for various initial conditions}
  %	\label{fig11}
  %\end{figure}
 A common conclusion arisen from the previous different plots in the case of dynamical system  (\ref{syst2}) for $m=0$ can be summarized as follow: although  radiation perfect fluid is added to the $f(T)$ gravity, there is always a global
 stable de Sitter attractor of all the cosmologies that satisfy $m\simeq0$. The corresponding stable $f(T)$   de Sitter attractor is asymptotically reached and reads $X^1_*=(0, -3,3,0)$. Without the $u$ coordinate, the equilibrium becomes the stable de Sitter attractor of the vacuum $f(T)$ gravity. However, the asymptotic behaviours shown by all plots in  the framework of  dynamical system (\ref{syst2}) for $m=0$ reveal that the variable $u\longrightarrow 0$  when $N\longrightarrow 60$. Regarding the expression of the variable $u$,  it follows that $u\longrightarrow 0$ implies $P_r\longrightarrow 0$ which means no radiation dominated. This is physically correct, since at a de Sitter point, neither the mass nor the radiation perfect fluids dominate the evolution \cite{Sergy}.
 
  \subsection{Matter dominated Attractors and their Stability in non-vacuum $f(T)$ gravity: $m= -9/2$}
 Let's search here for the cosmological evolution  when we consider the matter dominated era $m=-9/2$. We still come back to the dynamical system (\ref{syst2}) in which we replace the parameter $m$ by $-9/2$. The equilibria and their associated eigenvalue are presented in the table below 
 \begin{center}
 	\begin{tabular}{|c|c|c|c|c|c|c|c|c|}
 		\hline Fixed points & $x$  & $y$ & $z$ &$u$& $\eta_1$ &  $\eta_2$ &  $\eta_3$ &  $\eta_4$ \\ 
 		\hline $X^1_*$ & $-\frac{3}{2}$ & $-3$ & $\frac{3}{2}$& $0$ & $6$ & $-\frac{3}{2}$ & $ 0$& $\frac{3}{2}$ \\ 
 		\hline $X^2_*$  & $\frac{3}{2}$ & $-3$ & $\frac{9}{2}$ &$0$& $-6$ & $-\frac{9}{2}$ & $-\frac{3}{2}$&$0$ \\ 
 		\hline $X^3_*$ & $-3$ & $-\frac{9}{2}$ & $\frac{3}{2}$ &$1$& $6$ & $-\frac{3}{2}$ & $-\frac{7}{4}-i\frac{\sqrt{23}}{4}$& $-\frac{7}{4}-i\frac{\sqrt{23}}{4}$ \\ 
 		\hline $X^4_*$ & $3$& $-\frac{3}{2}$& $\frac{9}{2}$&$-3$& $-6$& $-\frac{3}{4}(3+\sqrt{17})$& $\frac{3}{4}(-3+\sqrt{17})$&$\frac{3}{2}$\\
 		\hline $X^5_*$ & $6$ & $\frac{3}{2}$ & $\frac{9}{2}$& $0$ & $-6$ & $3$ & $ \frac{9}{2}$& $\frac{9}{2}$ \\
 		\hline $X^6_*$  & $\frac{3}{2}$ & $\frac{3}{2}$ & $\frac{9}{2}$& $0$ & $-6$ & $-\frac{9}{2}$ & $ -\frac{3}{2}$& $\frac{9}{2}$ \\
 		\hline $X^7_*$ & $0$ & $-\frac{3}{2}$ & $\frac{3}{2}$& $0$ & $6$ & $\frac{3}{2}$ & $\frac{3}{2}$& $3$ \\
 		\hline $X^8_*$  & $-\frac{3}{2}$ & $-\frac{3}{2}$ & $\frac{3}{2}$& $0$ & $6$ & $-\frac{3}{2}$ & $ \frac{3}{2}$& $\frac{3}{2}$ \\
 		\hline 
 	\end{tabular} 
 \end{center}
 The system has eight different non-hyperbolic fixed points among which, only one is stable, $X^2_*=(\frac{3}{2},-3, \frac{9}{2},0)$. Once again, we   meet two values $z=\frac{3}{2}$ and $z=\frac{9}{2}$ which respectively, lead to $\omega_{eff}=1$ (matter dominated era) and $\omega_{eff}=-2$ (phantom evolution with dark energy as candidate). The behaviours of the dynamical  system solution can be visualised through the figure \ref{fig12}. 
  % \begin{figure}[h]
   %	\centering
   	%\begin{tabular}{rll}
   	%	\includegraphics[width=8cm, height=5cm]{cap23.eps}&
   	%	\includegraphics[width=8cm, height=5cm]{cap24.eps}
   	%\end{tabular}
   	%\caption{ The behaviour of    dynamical system (\ref{syst2}) solutions for $m=-9/2$ and for initial conditions $x(0)=-7$,$y(0)=-8$,$z(0)=5$ and $u(0)=4$}
   	%\label{fig12}
   %\end{figure}
 The figure
 % \ref{fig12}
  shows that the stable fixed point  $X^2_*$ is also reached asymptotically. The behaviours of the solutions around this  fixed point yields: when $N\longrightarrow60$,  one has $z\longrightarrow 9/2$ (phantom evolution or dark energy dominated era) and $u\longrightarrow 0$ (radiation perfect fluid is disappearing). This means that when  dark energy dominated era takes place, the radiation dominated era absconds.
 
 \section{Conclusion}
In this paper, we developed the modified Tele-Parallel $f(T)$ gravity  being a field of study of phase space  structure through the reconstruction and analysis of dynamical autonomous systems. The concerned dynamical systems are qualified autonomous because, in the building approach of the dynamical system, the  possible time-dependence or e-folding number dependence is recast in the parameter $m$ which is assumed to be constant along our investigation. Two different values with physical significance are considered during this investigation, namely, $m\simeq0$ which corresponds to de Sitter inflationary era and $m=-\frac{9}{2}$ linked to matter dominated era. After extracting  general dynamical systems from the $f(T)$ Friedmann motion equations, our analyses have taken place around two cosmological fundamental concepts: the vacuum $f(T)$ gravity  and the non-vacuum $f(T)$ gravity. In the framework of  the vacuum $f(T)$ gravity, the dynamical system has led to stable and unstable fixed point for both  $m\simeq0$ and  $m=-\frac{9}{2}$. But for $m\simeq0$, the numerical analyse of the dynamic system has yielded an asymptotic stable de Sitter attractor. Viable $f(T)$ models nearing this point and those unstable have been reconstructed. The matter dominated era in vacuum $f(T)$ gravity, after numerical investigation, has provided  stable attractor point which describes a phantom evolution like dark matter dominated era.  The second great and final step of the present work has also been based on the same previous analysis but now in the context of  non-vacuum $f(T)$ gravity. In this case and according to the used equation of motion, the radiation has been added to $f(T)$ gravity. Despite this addition, the numerical analysis for $m=0$ reveals an asymptotic stable de Sitter attractor around which the variable describing the radiation pressure vanishes. This result, also met in the case of the vacuum $f(T)$ gravity, holds physically because at the de Sitter point, neither the mass nor the radiation perfect fluids dominate the evolution. Finally, matter dominated era case in  non-vacuum  $f(T)$ gravity,  has given not only unstable fixed point but also  an  asymptotic  stable and probably attractor point which describes dark energy dominated era. At this phantom point, the  radiation variable goes asymptotically to zero.

 {\bf Acknowledgments}:

\begin{center}
 \rule{8cm}{1pt}
\end{center}

\end{document}